\documentstyle[12pt]{article}
\begin{document}

\def \ggww{\gamma\gamma\to WW}

\begin{center}
{\Large \bf
Strong Final State Interactions in $\ggww$}\\[2cm]
{\large P. Poulose\footnote{poulos@physik.rwth-aachen.de} 
and L.M. Sehgal\footnote{sehgal@physik.rwth-aachen.de}\\[5mm]
Institute for Theoretical Physics E\\
RWTH Aachen, D-52056 Aachen, Germany}
\end{center}

\vskip 4cm
\centerline{\bf Abstract}

\vskip 6mm
\noindent
We study the effects of a possible strong final state interaction 
among longitudinal $W$'s in $\ggww$. 
The relevant partial wave amplitudes are modified by an Omn\`es function 
approximated by a Breit-Wigner form factor.   We study the fractional 
cross section $f_{00}=\sigma_{LL}/\sigma_{\rm Total}$ with both $W$'s
longitudinally polarised, in the presence of a $J=2$ resonance ($M=2.5$
TeV, $\Gamma=500$ GeV) or a $J=0$ resonance ($M=1$ TeV, $\Gamma=1$ TeV),
whose parameters are scaled up from the $f_2$ and $\sigma$ resonances in
the $\pi\pi$ system.  We also examine the effects of final state interaction
in the case of polarised photons ($J_z=0$ or $J_z=2$), and the impact on
the Drell-Hearn-Gerasimov sum rule.

\newpage
\def \eeww{e^+e^-\to W^+W^-}
\def \ggvv{\gamma\gamma\to W^+W^-,\;ZZ}
\def \ggww{\gamma\gamma\to W^+W^-}

\section{Introduction}
The mechanism of 
electroweak symmetry breaking and the associated question of the origin
of masses of particles is a central issue in our understanding
of nature at a fundamental level. The Higgs mechanism employed in 
the standard model (SM) leads to the prediction of an elementary scalar 
particle with a mass $m_H=\sqrt{\lambda}\;v$, 
where $v=\left(\sqrt{2}\;G_F\right)^{-\frac{1}{2}}=246$ GeV, and $\lambda$ is
the quartic coupling of the scalar potential.  If the  coupling $\lambda$
is very large, however, the scalar sector becomes strongly interacting, and
the above relation for the Higgs mass breaks down.  In these circumstances,
even the existence of an elementary scalar particle becomes questionable.
Instead, the dynamics of the electroweak Goldstone bosons, which manifest
themselves as the longitudinal components of the gauge fields $W^\pm$ and
$Z$, begins to resemble the dynamics of the $\pi^\pm,\;\pi^0$ mesons, which
are the Goldstone bosons of chiral symmetry breaking in hadronic physics.
In such a situation, it is reasonable to expect resonances in the strong
interaction of longitudinal gauge bosons analogous to the
$\sigma,\;\rho,\;f_2,$ etc., in the $\pi\pi$ system. The possibility of
strong electroweak symmetry breaking is, therefore, best studied by
analysing processes involving $W_L$'s and $Z_L$'s.

Such a situation has been considered in the literature in 
various reactions. A $\rho$-like resonance in the electroweak sector
affects the process $\eeww$. Model independent analysis of this
process has been carried out by several authors \cite{eeww1}.
A specific model under the name of BESS has been developed to study
such new vector resonances \cite{bess}. While $\eeww$ is sensitive to 
the existence of a $\rho$-like resonance, possible
scalar and tensor resonances, which are the equivalents
of $\sigma$ and $f_2$ in the hadronic system, are conveniently studied 
in processes like $\ggvv$. Studies in this direction have been 
carried out in \cite{chanowitz}.

In our earlier study of $\eeww$, the strong final state interaction (FSI) 
of the longitudinal $W$'s was implemented with
the help of the BESS model, in which a $\rho$-like vector triplet arises
as a consequence of an additional $SU(2)$ gauge symmetry \cite{poul_eeww}. 
Such spin-1 resonances do not arise in $\gamma\gamma$ collisions.  
For our present investigation of $\ggww$, we take a model independent 
approach, exploring the consequences of $J=0$ and $J=2$ resonances in 
the $WW$ system, which modify particular partial waves in the amplitude.
These resonances are taken to be the equivalents of $\sigma$ and 
$f_2$ resonance in the $\pi\pi$ system.

The plan of the paper is as follows.
In Section \ref{sec-sm}, we recall the salient features of the process
$\ggww$ in the SM. 
In Section \ref{sec-fsi} we describe how strong final state interaction 
effects are included. In Section \ref{sec-result} we discuss
observables sensitive to such modifications. 
We give our conclusions in Section \ref{sec-conclusion}.

\def \non{\nonumber}
\def \be{\begin{eqnarray}}
\def \ee{\end{eqnarray}}
\def \la{\lambda}
\def \lga{\lambda_{\gamma_1}}
\def \lgb{\lambda_{\gamma_2}}
\def \lwp{\lambda_{w^+}}
\def \lwm{\lambda_{w^-}}

\def \eeww{e^+e^-\to W^+W^-}
\def \ggvv{\gamma\gamma\to W^+W^-,\;\;ZZ}
\def \ggww{\gamma\gamma\to W^+W^-}

\def \logbeta{\frac{L}{2\beta}}
\def \logbetaB{\frac{L}{\beta}}
\def \logbetaC{L}
\def \logbetaA{\log\left(\frac{1+\beta}{1-\beta}\right)}

\section{$\ggww$ in the Standard Model}

The process $\ggww$ in the Standard Model
has  $t$-channel and $u$-channel contributions
through the exchange of a $W$, and a contact interaction term.  
The helicity amplitudes are given in the literature \cite{ggww-sofar}, 
and we reproduce those in the Appendix.

Beam polarisation can be achieved in the photon linear colliders, 
where Compton back scattering of polarised laser and electron beams is used 
to produce high energy photon beams \cite{ph-collider}.
The two independent polarisation combinations of the photon beams
are those with same and opposite helicities. These are usually denoted by 
$J_z=\lga-\lgb=0$ and $J_z=2$, where $\lga$ and $\lgb$ are the helicities of the
two photons. Differential cross sections corresponding to these two cases
(summed over $W$ polarisations) are given below.

\begin{eqnarray}
\frac{d\sigma_{J_z=0}}{d\cos\theta} 
	&=& \frac{\beta}{8\pi s}\;
        \frac{(4\pi\alpha)^2}{(1-\beta^2\cos^2\theta)^2}\;
        \left(16-16r+3r^2\right)
	 \non\\[3mm]
\frac{d\sigma_{J_z=2}}{d\cos\theta} 
	&=& \frac{\beta}{8\pi s}\;
        \frac{(4\pi\alpha)^2}{(1-\beta^2\cos^2\theta)^2}\;
	\left[\left(3+10r+3r^2\right)\right.+ \non\\
        &&\left.2\left(5-2r-3r^2\right)\;\cos^2\theta+ 
        \left(3-6r+3r^2\right)\;\cos^4\theta\right]
        \non\\
\label{eq-diff-sig-sm}
\end{eqnarray}
Here, 
$\theta$ is the scattering angle, $\beta$ is the velocity of $W$
in the centre of mass frame, and $r=\frac{4m_w^2}{s}$. 
The unpolarised cross section is

\begin{eqnarray}
\frac{d\sigma}{d\cos\theta}
&=& \frac{1}{2}\;\left(\frac{d\sigma_{J_z=0}}{d\cos\theta}+
	\frac{d\sigma_{J_z=2}}{d\cos\theta}\right) \non\\[3mm]
&=& \frac{\beta}{16\pi s}\;
        \frac{(4\pi\alpha)^2}{(1-\beta^2\cos^2\theta)^2}\;
	\left\{\left(19-6r+6r^2\right)\right.+\non\\
        &&\left.
	2\left(5-2r-3r^2\right)\;\cos^2\theta+ 
       \left(3-6r+3r^2\right)\;\cos^4\theta\right\}
        \non\\
\end{eqnarray}
Both $J_z=0$ and $J_z=2$ distributions peak along the beam 
directions.
Integrated cross sections in the case of polarised beams have the
following expressions:

\begin{eqnarray}
\sigma_{J_z=0}(s)
        &=&\frac{\beta}{8\pi s}\;(4\pi\alpha)^2\;
        \left\{\left(\frac{16}{r}-16+3r\right)+
        \left(16-16r+3r^2\right)\;
        \logbeta \right\}     \non\\
\sigma_{J_z=2}(s)
        &=&\frac{\beta}{8\pi s}\;(4\pi\alpha)^2\;
        \left\{\left(\frac{16}{r}+22+3r\right)+
        \left(-16+4r+3r^2\right)\;
	\logbeta\right\} 
\label{eq-sigJz0Jz2}
\end{eqnarray}
where $L=\logbetaA$.

Notice that asymptotically these cross sections become equal, 
and the unpolarised cross section saturates to \(\sigma(s\to
\infty)=\frac{(4\pi\alpha)^2}{2\pi m_w^2}\sim \;80.9 \;\;{\rm pb}.\)
This advantage of $\ggww$ over other processes
like $\eeww$, which decrease with $s$, makes it
an attractive process in the linear colliders running at high energies.

The difference of the cross sections in Eq. \ref{eq-sigJz0Jz2}
\begin{eqnarray}
\Delta\sigma=\sigma_{J_z=2}-\sigma_{J_z=0}&=&
	\frac{\beta}{8\pi s}\;(4\pi\alpha)^2\;
	\left\{38+
	\left(-32+20\;r\right)\;
	\logbeta \right\}
\label{eq-dhg-ggww}
\end{eqnarray}
fulfills the sum rule
\begin{eqnarray}
\int_{\frac{4m_w^2}{s}}^\infty\frac{\Delta\sigma(s)}{s}\;ds = 0.
\label{eq-dhg}
\end{eqnarray}
This is a special case of the generalized Drell-Hearn-Gerasimov sum rule
\cite{dhg}, 
which applies to any elementary process $\gamma+a\to b+c$ in a gauge 
theory, $\Delta\sigma$ being the cross section difference for parallel and 
antiparallel  spins in the intial state.  It is worth noting that 
$\Delta\sigma$ changes sign at $\sqrt{s}=296$ GeV.  Any new physics in the
amplitude of $\ggww$ is likely to shift the location of this zero, and
affect the convergence of the sum rule.

Finally we quote the $\ggww$ cross section for longitudinal $W$'s. 
The angular distributions and integrated cross section for the production of
$W_LW_L$ are the following.

\begin{eqnarray}
\frac{d\sigma^{W_LW_L}_{J_z=0}}{d\cos\theta} 
	&=& \frac{\beta}{8\pi s}\;
        \frac{(4\pi\alpha)^2}{(1-\beta^2\cos^2\theta)^2}\;r^2
	 \non\\[3mm]
\frac{d\sigma^{W_LW_L}_{J_z=2}}{d\cos\theta} 
	&=& \frac{\beta}{8\pi s}\;
        \frac{(4\pi\alpha)^2}{(1-\beta^2\cos^2\theta)^2}\;
	(1+r)^2\;\sin^4\theta \non\\
\label{eq-diff-sig-sm-LL}
\end{eqnarray}
\begin{eqnarray}
\sigma^{W_LW_L}_{J_z=0}
&=&\frac{\beta}{8\pi s}\;(4\pi\alpha)^2\;
	\left[r+r^2\;\logbeta\right]\non\\
\sigma^{W_LW_L}_{J_z=2}
&=&\frac{\beta}{8\pi s}\;(4\pi\alpha)^2\;\frac{(1+r)^2}{\beta^2}\;\left[(2+r)
	+r\;(r-4)\;\logbeta\right] \non\\
\end{eqnarray}

Whereas the total cross section, summed over the $WW$ polarisations, goes
to a constant at high energy, the $W_LW_L$ fraction decreases with $s$.
This fraction is of particular interest as a probe of FSI involving
longitudinally polarised $W$'s.

In the following section we will discuss how the FSI is introduced by 
modifying particular partial waves, and how this affects various observables.

\label{sec-sm}

\section{Strong Final State Interactions}

As mentioned in the introduction, we assume the existence of strong interactions
among longitudinal gauge bosons, analogous to the case of the $\pi\pi$ system.
Final state interaction in the $\pi\pi$ system is known to introduce 
a phase shift in specific partial waves \cite{watson}, which 
may be parametrised in terms of an Omn\`es function $\Omega(s)$ 
\cite{omnes}. The partial wave amplitude in the presence of FSI is 
then modified according to 

\begin{eqnarray} 
M^J\to \Omega^J(s)\;M^J
\label{eq-ampl-modi-gen}
\end{eqnarray}

In the following we describe the partial wave decomposition of the
amplitude of the process $\gamma\gamma\to W_LW_L$, and how the 
FSI is introduced. The relevant helicity amplitudes have the expansion
\cite{martin}
 
\begin{eqnarray}
M^{SM}_{\lga\lgb 00}(s,\theta)&=&
\sum_J\frac{(2J+1)}{4\pi}\; {d}^{J\;*}_{J_z,0}(\theta)\; M_{\lga\lgb 00}^J(s)
\label{pwavedecomb}
\end{eqnarray}
Here ${d}^{J}_{J_z,0}(\theta)$ with $J_z=\lga-\lgb$ are 
the rotation functions corresponding to the helicities considered. 
 
Inverting Eq.\ref{pwavedecomb} using the orthogonality of the
rotation functions, we get the partial wave amplitudes
\begin{eqnarray}
M_{\lga\lgb 00}^J(s)=(2\pi)\;\int\!d\cos\theta\;
        {d}^{J}_{J_z,0}(\theta)\;M^{SM}_{\lga\lgb 00}(s,\theta) 
\label{pwaveampl}
\end{eqnarray}

As in the case of the $\pi\pi$ system \cite{morgan}, we expect the FSI 
to induce resonances of specific spins $J_R$. Such a resonance affects the 
partial wave amplitude $M^{J_R}_{\lga\lgb 00}$. In the case of $\gamma\gamma$ 
collisions, only even values of $J_R$ are allowed, and we will consider $J_R=0$ 
and $J_R=2$ resonances. In accordance with Eq. \ref{eq-ampl-modi-gen}, 
we introduce these effects through an Omn\`es function $\Omega(s)$.  
For simplicity, we approximate this function by a Breit-Wigner 
fuction normalised to unity at the threshold. That is,

\begin{eqnarray}
M_{\lga\lgb 00}^{J_R}(s)\to
        \Omega^{J_R}(s)\;M_{\lga\lgb 00}^{J_R}(s),
\end{eqnarray}
with
\begin{eqnarray}
\Omega^{J_R}(s) = \frac{4m_w^2-m_R^2}{s-m_R^2+i\;\Gamma_Rm_R\;
        \left(\frac{\beta}{\beta_R}\right)^{(2J_R+1)}}
\label{omega}
\end{eqnarray}
Here, $m_R$ is the mass and $\Gamma_R$ the width of the resonance,
$\beta=\sqrt{1-4m_w^2/s}$ and $\beta_R=\sqrt{1-4m_w^2/m_R^2}$.
This modifies the helicity amplitudes such that 
 
\begin{eqnarray}
M_{\lga\lgb 00}(s,\theta)&=&
\sum_J\frac{(2J+1)}{4\pi}\;
{d}^{J\;*}_{J_z,0}(\theta)\;
M_{\lga\lgb 00}^J(s) \non\\
&=&M^{SM}_{\lga\lgb 00}(s,\theta)+
        \left[\Omega^{J_R}(s) - 1\right]\;\frac{2J_R+1}{4\pi}\;
        d^{J_R\;*}_{J_z,0}(\theta)\;
        M_{\lga\lgb 00}^{J_R} (s)  \non\\
\label{pwavedecomb2}
\end{eqnarray}
where the superscript $SM$ denotes the standard model result. 

Expressions for the relevant $d^{J_R}_{J_z,0}$ functions and partial wave 
amplitudes are given in the Appendix. With the modified helicity 
amplitudes in Eq. \ref{pwavedecomb2}, differential cross 
sections given in Eq. \ref{eq-diff-sig-sm} are changed to 
 
\begin{eqnarray}
\left.\frac{d\sigma_{J_z=0}}{d\cos\theta}\right|_{J_R=0}
&=&\frac{d\sigma_{J_z=0}^{SM}}{d\cos\theta}+
	\frac{\beta}{32\pi s}\;(4\pi\alpha)^2\;\non\\
	&&\times\left\{
	2\;{\rm Re}(\Omega-1)\;
	\frac{2\;r^2\;\logbetaC}{\beta\;(1-\beta^2\;\cos^2\theta)} + 
        |\Omega-1|^2 \;\frac{r^2\;\logbetaC^2}{\beta^2} 
	\right\} \non\\[3mm]
\left.\frac{d\sigma_{J_z=0}}{d\cos\theta}\right|_{J_R=2}
&=&\frac{d\sigma_{J_z=0}^{SM}}{d\cos\theta}+
	\frac{\beta}{32\pi s}\;(4\pi\alpha)^2\;\non\\
	&&\times\left\{
	2\;{\rm Re}(\Omega-1)\;
	\frac{5}{\beta^2}\;2\;r^2\;
	\left[-3+(2+r)\;\logbeta\right]\;
	\frac{3\;\cos^2\theta-1}{1-\beta^2\;\cos^2\theta}+ \right.\non\\
        &&\left.\;\;\;\;\;
        |\Omega-1|^2 \;\left(\frac{5}{\beta^2}\right)^2\;r^2\;
	\left[-3+(2+r)\;\logbeta\right]^2\;
	\left(3\;\cos^2\theta-1\right)^2\right\} \non\\[3mm]
\left.\frac{d\sigma_{J_z=2}}{d\cos\theta}\right|_{J_R=2}
&=&\frac{d\sigma_{J_z=2}^{SM}}{d\cos\theta}+
	\frac{\beta}{32\pi s}\;(4\pi\alpha)^2\;\non\\
        &&\times\left\{
	2\;{\rm Re}(\Omega-1)\;
	\frac{5}{2\beta^4}\;(1+r)^2\;
	\left[(2-5\;r)+3r^2\;\logbeta\right]\;
	\frac{\sin^4\theta}{1-\beta^2\;\cos^2\theta}+\right. \non\\
        &&\left.\;\;\;\;\; \;\;\;
        |\Omega-1|^2 \;\left(\frac{5}{4\beta^4}\right)^2\;
	(1+r)^2\;
	\left[(2-5\;r)+3r^2\;\logbeta\right]^2\;
	\sin^4\theta\right\} \non\\
\end{eqnarray}

For convenience we have dropped the labels on $\Omega$.
The expressions for $\Omega$ in the case of 
$J_R=0$ and $J_R=2$ differ according to Eq. \ref{omega}. 
Corresponding integrated cross sections are given by 
\begin{eqnarray}
\left.\sigma_{J_z=0}(s)\right|_{J_R=0}
        &=&\sigma^{SM}_{J_z=0}(s)+
	\left(2\;{\rm Re}(\Omega-1)+|\Omega-1|^2\right)
	\;\frac{\left(4\pi\alpha\right)^2}{16\pi s}\;
	\frac{r^2}{\beta}\;
	\logbetaC^2 \non\\[3mm]
\left.\sigma_{J_z=0}(s)\right|_{J_R=2}
        &=&\sigma^{SM}_{J_z=0}(s)+
	\left(2\;{\rm Re}(\Omega-1)+|\Omega-1|^2\right)
        \non\\
        &&\times  \frac{\left(4\pi\alpha\right)^2}{16\pi s}\;
	\frac{5\;r^2}{\beta^3}\;
        \left[-3+(2+r)\;\logbeta\right]^2\non\\[3mm]
\left.\sigma_{J_z=2}(s)\right|_{J_R=2}
        &=&\sigma^{SM}_{J_z=2}(s)+
	\left(2\;{\rm Re}(\Omega-1)+|\Omega-1|^2\right)
        \non\\
        &&\times \frac{\left(4\pi\alpha\right)^2}{16\pi s}\;
	\frac{5}{6}\;\frac{\left(1+r\right)^2}{\beta^7}\;
        \left[\left(2-5\;r\right)+ 3\;r^2\;\logbeta
	\right]^2 \non\\
\label{sigmaModified}
\end{eqnarray}

The above equations are used to analyse how various cross
sections are modified.
We discuss the numerical results in the next section.

\label{sec-fsi}
 
\section{Numerical Results}
In our numerical analysis we have considered spin-0 and
spin-2 resonances analogous to $\sigma$ and $f_2$ in the 
$\pi\pi$ system, scaled up to $m_R=1$ TeV and $\Gamma_R=1$ TeV;
and $m_R=2.5$ TeV and $\Gamma_R=0.5$ GeV respectively,
using a rough scaling factor of $v/f_\pi\sim 2000$, where $f_\pi$ is the
$\pi$ decay constant. We keep in 
mind a photon-collider operating at and above 500 GeV centre of 
mass energy, as is expected, for example in
TESLA.  We also consider the polarisation option, which is 
expected to be achieved in photon-linear colliders.
Our principal results are as follows.

\begin{figure}[h]
\vskip 6cm
\includegraphics{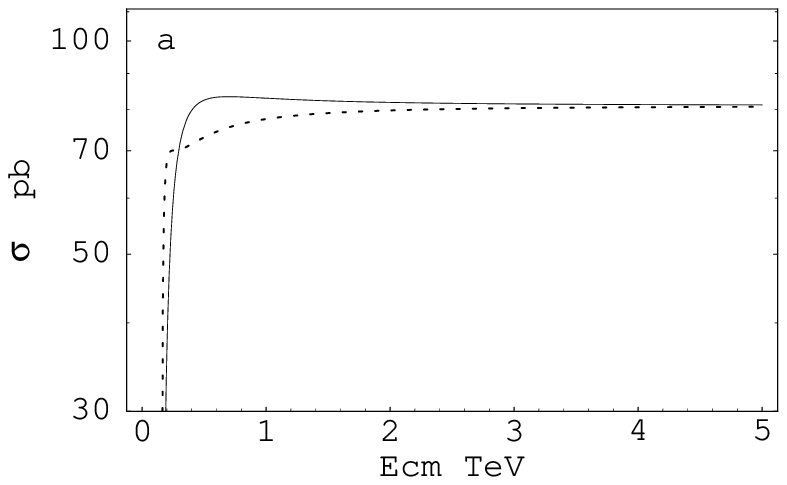}
\includegraphics{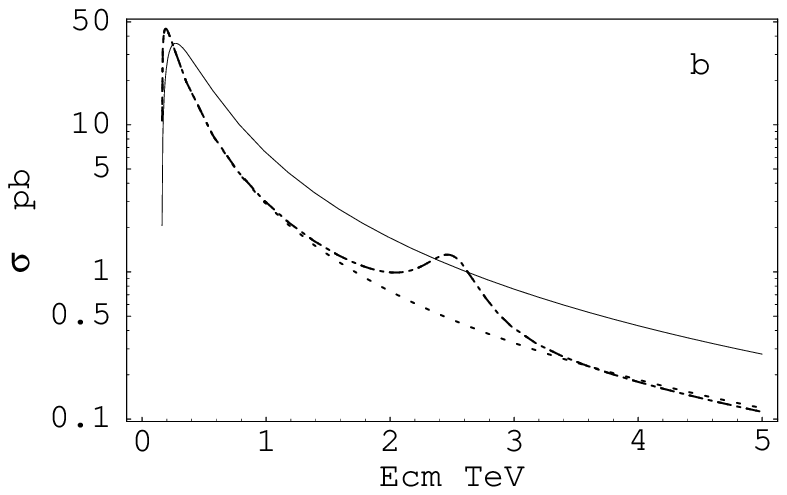}
\caption{Cross section of unpolarised $W$'s ($a$) without any cut, and
($b$) with an angular cut of $|\cos\theta|\leq 0.8$. Solid and dotted curves
correspond to $J_z=2$ and $J_z=0$ respectively, in the SM. The dash-dotted curve 
in ($b$) shows the effect of a $J_R=2$ final state resonance.
}
\label{fig-sig-unpol}
\end{figure}

{\large \it 1.} Fig. \ref{fig-sig-unpol}$a$
shows cross sections versus $\sqrt{s}$ separately for the polarisation 
states $J_z=0$ and $J_z=2$.  The effects of the resonances are essentially
invisible on this scale.  To increase the sensitivity to FSI, we consider
the cross section with an angular cut $|\cos\theta|\leq 0.8$.  This,
for the case of $J_R=2$ is shown in  Fig. \ref{fig-sig-unpol}$b$, 
where the resonant enhancement is visible. For the $J_R=0$ resonance,
the enhancement is essentially invisible even after the angular cut.

\begin{figure}[h]
\vskip 12cm
\includegraphics{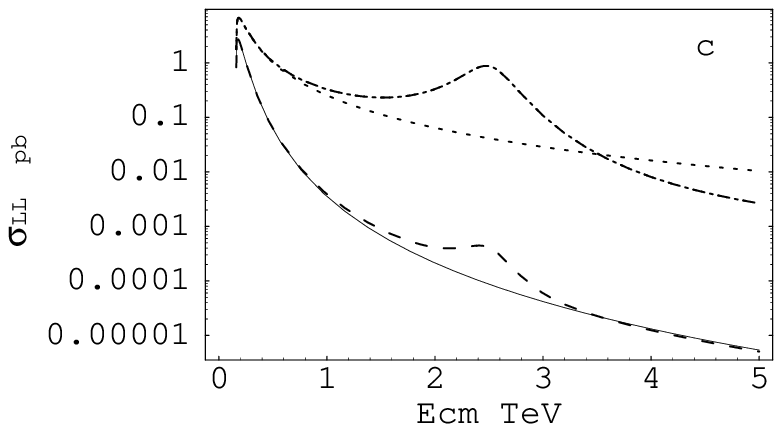}
\includegraphics{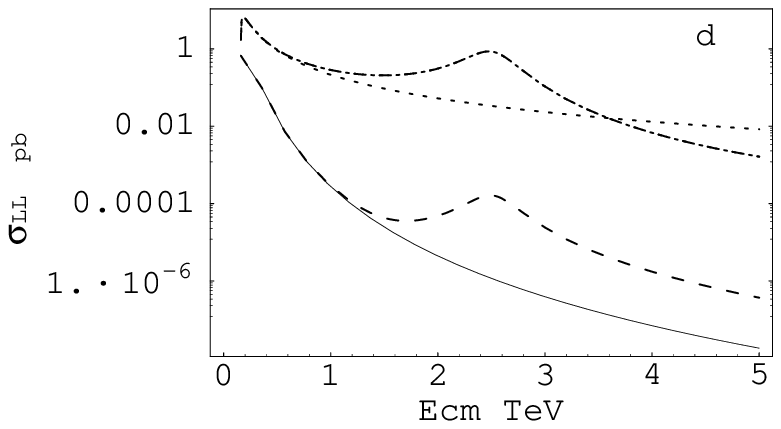}
\includegraphics{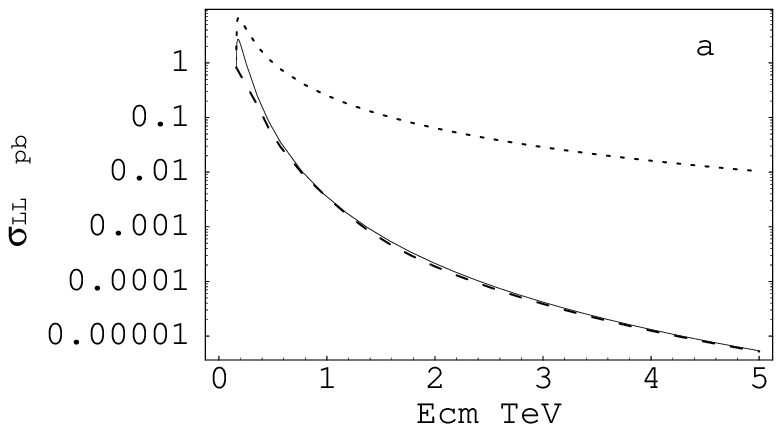}
\includegraphics{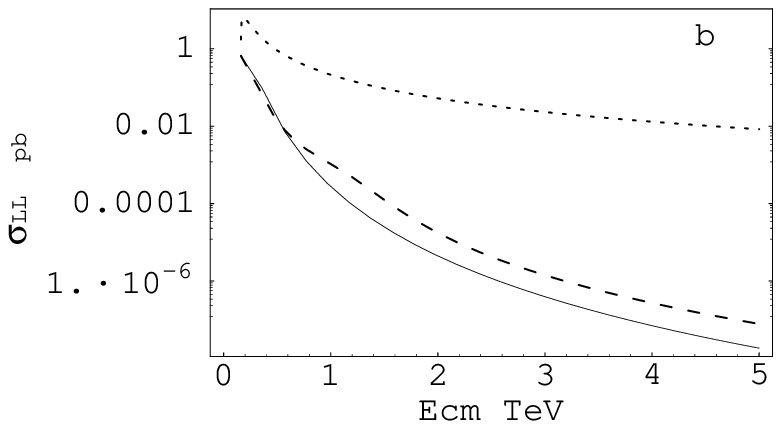}
\caption{
Cross section for longitudinally polarised $W$'s. ($a$) and ($c$) are without 
angular cut, while ($b$) and ($d$) are for $|\cos\theta|\leq 0.8$. Effects of 
$J_R=0$ are shown in ($a$) and ($b$), and for $J_R=2$ in ($c$) and ($d$).
Solid and dotted lines correspond to the SM results for $J_z=0$ and $J_z=2$ 
respectively, while the dashed and dash-dotted curves include FSI.
}
\label{fig-sig-WL}
\end{figure}

{\large \it 2.} In Fig. \ref{fig-sig-WL}$a$ we show a plot of $\sigma_{LL}$ 
(the cross section for $W_LW_L$) versus
$\sqrt{s}$, with and without the $J=0$ and $J=2$ resonances. The effects of
an angular cut are indicated in Fig. \ref{fig-sig-WL}$b$.
Clearly, the effects of final state interaction become more visible when the 
longitudinal part of the coss section ($\sigma_{LL}$) is isolated.

{\large \it 3.} The longitudinal fraction of the cross section, defined as
\[ f_{00} = \frac{\sigma_{LL}}{\sigma_{\rm tot}}, \]
is given in Table 1.  Values of $f_{00}$ are listed
in the SM and in the presence of FSI with and without an angular cut.  One
can observe that, with an angular cut $|\cos\theta|\leq 0.8$,
the fraction $f_{00}$ increases typically from 3\% in the SM to 4\% 
in the presence of  FSI. 
It may be recalled that the parameter $f_{00}$ can be
determined empirically by studying the energy distribution of the secondary
leptons from the $W$'s (see Ref.\cite{anja}).

\begin{table}[h]
\begin{center}
\begin{tabular}{|l|l|c|c|c||c|c|c|}
\hline
\multicolumn{2}{|c|}{}&\multicolumn{3}{|c||}{}&\multicolumn{3}{|c|}{}  \\
\multicolumn{2}{|c|}{}&\multicolumn{3}{|c||}{No angular cut}&
	\multicolumn{3}{|c|}{$|\cos\theta|\leq 0.8$}\\[3mm]\cline{3-8} 
\multicolumn{2}{|c|}{}&&&&&& \\ 
\multicolumn{2}{|r|}
{$\sqrt{s}$ (GeV)}&500&800&1000&500&800&1000\\[3mm]\cline{1-8}
\multicolumn{2}{|c|}{}&&&&&& \\ 
\multicolumn{2}{|r|}{$\sigma^{SM}$ (pb)}&
77.82& 79.90& 80.32& 15.75& 7.05& 4.66\\[3mm]\cline{1-2}
&{\small SM}&0.0068&0.0025&0.0016&0.029&0.024&0.023\\
$f_{00}$&{\small SM+FSI($J_R=0$)}&0.0122&0.0036&0.0021&0.045&0.030&0.027\\
&{\small SM+FSI($J_R=2$)}&0.0125&0.0040&0.0026&0.046&0.034&0.034\\
[3mm]\cline{1-8}
\end{tabular}
\caption{Longitudinal fraction of cross section, $f_{00}=\sigma_{LL}/\sigma_{\rm tot}$ 
with and without angular cut. Values of $f_{00}$ in the presence of
FSI with $J_R=0$ and $J_R=2$ resonances are shown along with the SM values, for 
different centre of mass energies of the collider.}
\end{center}
\label{table-f00}
\end{table}

\begin{figure}[h]
\vskip 5cm
\includegraphics{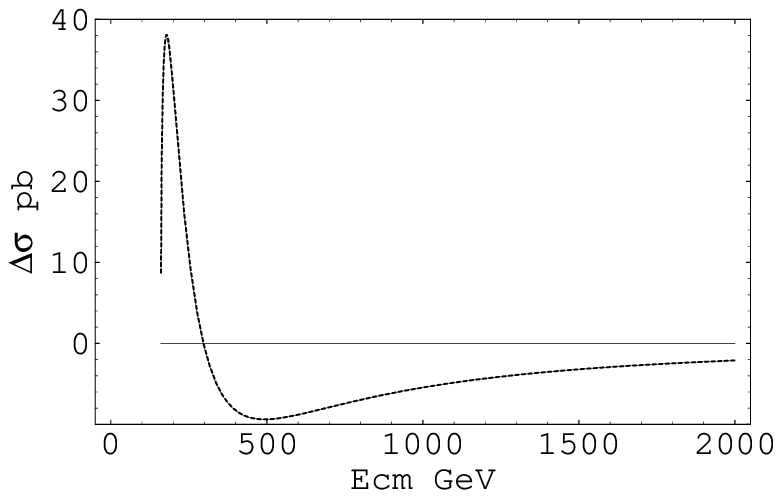}
\includegraphics{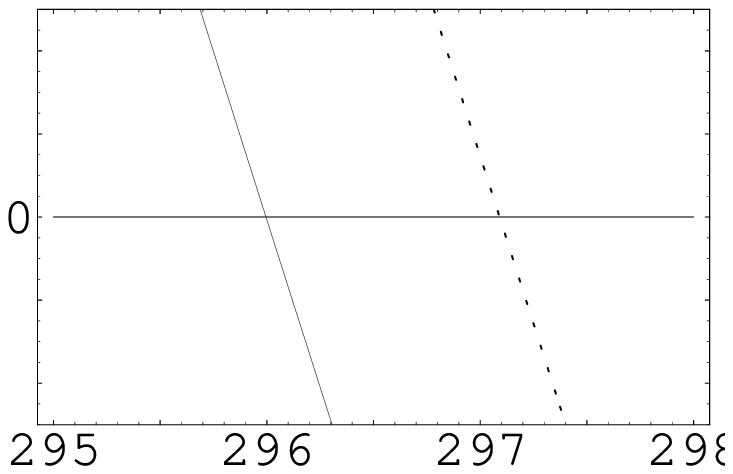}
\caption{${\Delta\sigma}$ against $\sqrt{s}$. Inset shows the 
shift in the position of zero (dotted line) in the presence of strong FSI with 
$J_R=0$ resonance at $m_R=1$ TeV and $\Gamma_R=1$ TeV. }
\label{fig-dhg}
\end{figure}

{\large \it 4.} Finally, we show in Fig. \ref{fig-dhg} the cross section difference
$\Delta\sigma=\sigma_{J_z=2}-\sigma_{J_z=0}$ as a function of $\sqrt{s}$.
The zero of the function, which in the SM lies at $\sqrt{s}=296$ GeV, is
shifted upwards by about 1 GeV. The integral (Eq. \ref{eq-dhg}) is no longer 
convergent, as is to be expected when gauge theory amplitudes are modified by 
an unconventional $WW$ interaction introduced in a phenomenological way.

\label{sec-result}

\section{Conclusions}
We have investigated the effects of strong $W^+_LW^-_L$ interaction 
in the channel $\ggww$, assuming resonant structures analogous to the
$\sigma$ and $f_2$ resonances in the $\pi\pi$ system.  Numerical
corrections to the standard model predictions have been evaluated for the
total cross section, and for the longitudinal fraction,
$f_{00}=\sigma_{LL}/\sigma_{\rm tot}$.  A typical result is an enhancement
of $f_{00}$ from 3\% to about 4\% at $\sqrt{s}=500$ GeV, after inclusion of
an angular cut $|\cos\theta|\leq 0.8$.  We have also calculated the cross
section difference, $\Delta\sigma=\sigma_{J_z=2}-\sigma_{J_z=0}$ for like 
and unlike photon polarisations in the initial state, which is relevant to
the generalized DHG sum rule.  The zero of $\Delta\sigma(s)$, which occurs
at 296 GeV in the SM, is shifted upwards by about 1 GeV in the presence of
FSI.
\label{sec-conclusion}

\vskip 1cm
\noindent
{\large \bf Acknowledgements}\\[2mm]
One of us, P.P. wishes to thank the Humboldt Foundation for a
Post-doctoral Fellowship, and the
Institute of Theoretical Physics E, RWTH Aachen for the
hospitality provided during this work.

\def \logbeta{\log\left(\frac{1+\beta}{1-\beta}\right)}

\vskip 10mm
\noindent
{\Large \bf Appendix A: Helicity Amplitudes}\\[5mm]
Following are the
helicity amplitudes of the process $\ggww$ in the Standard Model,
norlamised such that,
\begin{eqnarray}
\frac{d\sigma}{d\cos\theta} = \frac{\beta}{32\pi s}\;
                \sum_{\lambda} |M^{SM}_{(\lga,\lgb,\lwm,\lwp)}(s,\cos\theta)|^2
\end{eqnarray}
with $\lga,\lgb,\lwm$ and $\lwp$ are 
helicities of initial photon beams and $W$'s produced; $\sqrt{s}$, the
centre of mass energy; $\theta$ the scattering angle in the c.m.f.
Summation over final particle helicities, and averaging over initial
state helicites are implied by $\sum_\lambda$.
 
Taking out a common factor, the helicity amplitudes are expressed in
terms of the functions $A_{(\lga,\lgb,\lwm,\lwp)}$ as

\begin{eqnarray}
M^{SM}_{(\lga,\lgb,\lwm,\lwp)}
(s,\cos\theta)=\frac{4\pi\alpha}{1-\beta^2\cos^2\theta}\;
A^{SM}_{(\lga,\lgb,\lwm,\lwp)}(s,\cos\theta),
\end{eqnarray}

with
\begin{eqnarray}
A^{SM}_{\pm\pm 00} &=& 2\;r \non\\
A^{SM}_{\pm\pm\pm\pm} &=& 2\;(1+\beta)^2 \non\\
A^{SM}_{\pm\pm\mp\mp} &=& 2\;(1-\beta)^2 \non\\[5mm]
A^{SM}_{\pm\mp 00} &=&-2\;\left(1+r\right)\;\sin^2\theta \non\\
A^{SM}_{\pm\mp 0\pm} &=&\mp\;2\sqrt{2\;r}\;
		(1-\cos\theta)\;\sin\theta 
	=A^{SM}_{\mp\pm\pm 0} \non\\
A^{SM}_{\pm\mp 0\mp} &=& \mp\;2\sqrt{2\;r}\;
		(1+\cos\theta)\;\sin\theta 
	=A^{SM}_{\mp\pm\mp 0}\non\\
A^{SM}_{\pm\mp\pm\pm} &=& 2\;r\;\sin^2\theta
	=A^{SM}_{\mp\pm\pm\pm} \non\\
A^{SM}_{\pm\mp\pm\mp} &=& 2\;(1+\cos\theta)^2 \non\\
A^{SM}_{\pm\mp\mp\pm} &=& 2\;(1-\cos\theta)^2 \non\\
\end{eqnarray}
where $r=\frac{4m_w^2}{s}$. All other helicity amplitudes vanish.

\vskip 10mm
\noindent
{\large \bf Appendix B: Partial Wave Amplitudes}\\[5mm]

Partial wave amplitudes are given in terms of 
the helicity amplitudes as,
\begin{eqnarray}
M_{\lga,\lgb,\lwm,\lwp}^J(s)=(2\pi)\;\int\!d\cos\theta\;
        {d}^{J}_{\delta_\gamma,\delta_w}(\theta)\;
	M^{SM}_{\lga,\lgb,\lwm,\lwp}(s,\theta) 
\end{eqnarray}

Here $\delta_\gamma=\lga-\lgb$ and $\delta_w=\lwm-\lwp$. 
Partial wave amplitudes that are relevant in the present case
are the following:
 
\begin{eqnarray}
M_{\pm\pm 00}^{J_R=0} (s)
        &=&4\pi\alpha\;\frac{4\pi\:r}{\beta}\;\logbeta\non\\
M_{\pm\pm 00}^{J_R=2} (s)
        &=&4\pi\alpha\;
        \frac{4\pi\;r}{\beta^2}\;\left[-3+\frac{2+r}{2\beta}\;
        \logbeta\right]          \non\\[5mm]
M_{\pm\mp 00}^{J_R=2} (s)
        &=&-4\pi\alpha\;
        \frac{4\pi\;(1+r)}{\sqrt{6}\;\beta^4}\;
        \left[\left(2-5\;r\right)+\frac{3\;r^2}{2\beta}\;
        \logbeta\right]  \non\\
\label{partialwavesJ2}
\end{eqnarray}

We have used the following $d$ functions:
\begin{eqnarray}
d^{0}_{0,0} &=&1\non\\
d^{2}_{0,0} &=&\frac{1}{2}\;(3\;\cos^2\theta-1)\non\\
d^{2}_{2,0} &=&\sqrt{\frac{3}{8}}\;\sin^2\theta=d^{2}_{-2,0} 
\end{eqnarray}

\end{document}